\documentclass[12pt]{article}
\begin{document}

                                PARTITION FUNCTIONS FOR STATISTICAL MECHANICS 

 WITH MICROPARTITIONS AND PHASE TRANSITIONS

                                                               AJAY PATWARDHAN

                                          Physics department, St Xavier's college,

 Mahapalika Marg , Mumbai 400001,India

                                            Visitor, Institute of Mathematical Sciences, 

 CIT campus, Tharamani, Chennai, India

                                                           ajay@imsc.res.in

                                                             ABSTRACT

The fundamentals of Statistical Mechanics require a fresh definition in the context of the developments in Classical Mechanics of integrable and chaotic systems. This is done with the introduction of Micro partitions; a union of disjoint components in phase space. Partition functions including the invariants ,Kolmogorov entropy and Euler number are introduced. The ergodic hypothesis for partial ergodicity is discussed.

In the context of Quantum Mechanics the presence of symmetry groups with irreducible representations gives rise to degenerate and non degenerate spectrum for the Hamiltonian. Quantum Statistical Mechanics is formulated including these two cases ; by including the multiplicity dimension of the group representation and the Casimir invariants into the Partition function. The possibility of new kinds of phase transitions is discussed.

The occurence of systems with non simply connected configuration spaces and Quantum Mechanics for them , also requires a  possible generalisation of Statistical Mechanics. The developments of Quantum pure, mixed and entangled states has made it neccessary to understand the Statistical Mechanics of the multipartite N particle system . And to obtain in terms of the density matrices , written in energy basis , the Trace of the Gibbs operator as the Partition function and use it to get statistical averages of operators. There are some issues of definition, interpretation and application that are discussed.

 INTRODUCTION

A  Statistical Mechanics of systems with properties of symmetry , chaos, topology and mixtures is required for a variety of physics ; including condensed matter and high energy physics. When transformations involving changes in these properties occurs , a general framework of statistical mechanics is required. In defining ensembles ; microcanonical, canonical, grand canonical, and the partition functions to calculate the ensemble averages and thermodynamics, phase transitions of new kinds and explanations of phase transitions of  usual kinds are found. 

For classical phase space the partition function is defined , including the micro partitions or ensembles or configurations on constant energy surfaces. These include the additional indexes due to ergodic and topological components. For quantum systems the density matrix form is generalised to get a mixed system of pure and entangled states and symmetry group representation states are used.

These constructions show that there are more underlying structures possible in statistical physics. The canonical ensemble is the central one , which has a precursor as micro canonical and a extension to grand canonical ensemble. Some configurations in micro canonical ensemble acquire more possibilities , while in the grand canonical ensemble the sum over the numbers of particles is also altered ,  as the structures of the underlying system can be dependent on the number of particles.

$(1)$ CLASSICAL STATISTICAL MECHANICS WITH INTEGRABLE KAM TORII AND CHAOTIC REGIONS IN PHASE SPACE

For a N particle system the $ 6N $ dimensional classical phase space is a manifold $ M $ on which Hamiltonian dynamics is defined. The invariant sub manifolds of fixed energy are given by the K A M torii (integrable regions or islands ) and ergodic components ( or chaotic seas) . These submanifolds partition the phase space into disjoint subspaces with specific invariants. This may be called partial ergodicity or broken ergodicity.

 For Statistical Mechanics the ergodic hypothesis of phase space averages equal to long time averages , has the need for invariant ergodic measures,  to be defined on the union of these subspaces , with characteristic set functions. As the integration measure has multiplicative property and the characterising invariants have an additive property ; the integration measure with a weight function that is exponential is required , with the characterising quantity in the exponent. This quantity is the Kolmogorov Entropy for the chaotic seas and the topological invariant Euler number for the islands .

 The critical points of the hamiltonian vector field and the Hessian ( determinant of the second partial derivatives of the Hamiltonian ) give the local conditions for the K A M torii. The Euler number invariant is a sum of the index at the  singular points of the vector field . Hence it is proposed that these two quantities which are invariant on the submanifolds be taken as the exponents in a weight function . For the closed ,integrable KAM torii subspaces $M$, the Hamiltonian vector field has isolated and finite number  $ 'p'$ singularities . The index of the vector field $'v'$ at the singular point $'p'$ is $J_v(p) $, then the sum $\sum J_v(p)  = \chi(M) $ is the Euler characteristic of the subspace $ M $. 

With these micro partitions to use in the phase space averages,  the ergodic hypothesis is restored in the presence of distinct regions of ergodicity. Since all possible initial conditions in phase space are now averaged over , the canonical partition of constant energy hypersurfaces have the micro partition of these integrable and ergodic regions . For all initial conditions in a particular phase space component the dynamics takes the system to every point either in a ergodic invariant (chaotic) set or a dense invariant set (integrable KAM torii). This is a transitive property. The long time average then becomes equal to the phase space average when all the micro partitions are integrated and summed over.

 The probability of visiting each phase space volume depends on a invariant measure associated to it , as defined below. Hence a large stochasticity of trajectories signified by a a larger Kolmogorov entropy ,which is an invariant on each distinct ergodic component, gives a higher weightage function for the chaotic  component. Also a higher value of the Euler index , which is an invariant for each distinct integrable component, gives a higher weightage to the integrable component. Thus the uniform distribution of the present form of statistical mechanics is replaced by the weighted one in this generalisation.

 If only one ergodic component  in the whole phase space is obtained, as it is when the energy is largre, or the chaos parameter is beyond a critical value ,or the number of particles is large, then the formalism reduces to the usual statistical mechanics . Because now the integrable regions occupy negligible phase space volume and the Euler number tends to zero; as well as the Kolmogorov entropy is absorbed into the fiducial entropy of the system and contributes a fixed constant number in the integral in the limit of large energy and number of particles.

The Partition function is written as 

$ Z(M) = \sum_{M_n}\int Exp (-\beta H) Exp (\beta\sum_i \mu _iC_i)\delta(H(M_n) -E) m(M_n) dM_n$ 

Where the invariant weight function  $ m(M_n) $  on the disjoint components $ M_n $  is given by the $ Exp(K(E)_n) $ for the chaotic part

 and  $ dim(M_n) ^{\chi(M_n)} $ for the integrable  KAM torii parts. These micro partitions $ M_n $ are governed by the value of 

the Hessian $ det (\partial^2 H(M)/(\partial q \partial p) )$ being positive or negative or zero.

 $ H $ is the Hamiltonian on phase space $ M $ and $ K (E)_n $ is the Kolmogorov entropy , the sum of all positive Lyapunov exponents and it is  non zero for the chaotic component  $ M_n $ . The  $ \chi (M_n) $ is the Euler number index for the integrable KAM torii component $ M_n $ ; given by the sum of the critical points of the Hamiltonian vector field :  $ \beta $ will be $( kT)^{-1} $. The $\delta $ function in the integral is the microcanonical measure for the constant energy subspace. 

The presence of any conserved quantities $ C_i $ introduces the associated chemical potentials $\mu_i$ and hence a exponential factor in the grand canonical ensemble. $ dM_n $ is the standard integration measure on phase space $M$ restricted to the micro partition $ M_n$. The sum is over all the distinct ergodic  components denoted by $ M_n $.

The statistical average of a function on phase space , such as the Hamiltonian , is now obtained by taking this function and inserting it  into the integrand defined above for the partition function. Then the ratio of this functions integral and the partition function integral is the average value of the function.

 As an example consider a collection of composite particles whose internal Hamiltonian is given by the Henon Heiles Hamiltonian. This is a well known system in classical and quantum mechanics and for which the invariants are known. The expression for the canonical partition function can be defined with these quantities. The kolmogorov entropy for this system increases with energy and reaches an asymptotic value. And the volume fraction of the chaotic region increases, while the volume fraction of the integrable region in phase space decreases linearly with energy between two critical energies of the system. The Euler characteristic for Torii, with genus one, is zero; and for homotopically deformed KAM Torii it is computed from the critical points of the Hamiltonian vector field .  The Hessian condition for the micro partition into chaotic and integrable regions is computable as a function of energy. With numerical , if not analytic expressions available for these quantities ; the integrals in the partition function can be evaluated.

$(2)$ STATISTICAL MECHANICS WITH MULTIPLY CONNECTED CONFIGURATION SPACE

 Quantum Mechanics with multiply connected configuration space requires a Hilbert space with a fundamental group and higher homotopy groups representations. The statistical mechanics of the N particle system consists of density matrices that are partitioned into the direct sum of the basis representation. 

The equivalence classes of such representation are used to describe the types of phase transitions when a change in the equivalence class occurs. The union of all the disjoint subspaces is taken as the configuration space.

  A lattice space group in $R^n$, modulo a finite group ,  gives rise to topologically and geometrically interesting subspaces. This construction can give orientable spaces such as torii as well as non orientable ones such as Klein bottles. It can include the polyhedra of the space groups.

 The partition functions on such subpaces and their relation to phase transitions induced by a change of the modulii groups is a topic of relevance to field theory, string /brane theory and gravitational systems with horizons . Besides the more common occurence in condensed matter physics of topological groups ,  classification of phases and defects.

 These are reported in the research papers for some specific examples. This subject is still in need of a general framework , possibly a path integral/partition function approach. One example is the particles in an annular disc region. The presence of excluded points or subsets in a configuration space gives a ``swiss cheese'' topology . The fundamental and higher homotopy groups provide a representation on a direct sum of Hilbert space of states. The path integral over homotopically equivalent paths gives the kernels and their sum to get the Partition function. There are compilations in Handbook of path integrals for a number of problems.

The spacetime manifold has a 'swiss cheese' topology when a number of black holes with null hypersurfaces as horizons occur . The exterior solution has particle and field dynamics. The union of all the interiors of the horizons is the excluded region. For average seperations between two black holes as larger than 1000 time the horizon radius , the spacetime is almost flat between them. Path integral calculations for particle and field partition functions are possible. The problem is analogous to that in condensed matter with a dilute set of impurities in a background lattice. The excluded regions are the small neighbourhoods of each impurity. Then the path integral based partition function will have a sum over the equivalence classes of paths according to Homotopy. Phase transitions due to changes in the homotopy group for the multiply connected regions would correspond to merger of black holes in space time or extinction and creation of defects and  impurities in condensed matter.

 Thermodynamics of black holes is a important topic in gravitation and for multiple black holes , there are a few results. The collapse of the star's surface below the null surface of the horizon can continue till the numbers of particles occupy the volumes $h^3$ each in phase space. The entropy is then related to the number of configurations in this volume of quantum spacetime foam. Partitioning the volume into cells, as solid angles each with one particle configuration and summing over them , gives the area of the horizon  ; if the solid angles were projected up from the center to the horizon surface  of the 2 sphere. Better calculation methods may evolve by using the concept of micro partitions and partition functions in the multiply connected space times of a collection of black holes.; to understand the thermodynamics of the formation of black hole clusters and super massive black holes.

The vaccuum expectation value is $ <0\mid0> = Z = \int D\phi Exp(-S(\phi)) $ 

and the 'n ' point correlation function  with $ Z(J) = \int D\phi Exp(-S(\phi) +J\phi) $ 

is given by $\partial^n Z /\partial J^n $.

In this representation of the partition function the expansion of $Z(J) $ in terms of $Z(0) $

requires the expansion of the Action $ S[\phi] = S[\phi_0] + $ Det $ [ \partial^2 S/(\partial \phi_i \partial \phi_j )] $.

This involves the Van Vleck determinant which gives rise to the equations of motion or field equations , with the Green's function as the ``inverse'' of it.

The insertion of this Action into the partition function then gives a Gaussian integral , whose value is $[Det]^{-1/2}$.

The association of the $exp(-itH) $ to the $exp(-\beta H)$, and the Euclidean time and $\beta$ association, allows writing the resultant partition function in the alternate form

$ Z(\beta) = \int D[\phi] Exp(-S[\phi]) $ and $ Z(\beta) = $ Tr $ Exp(-\beta H)$.

For the homotopically distinct micro partitions to be summed over , there is need to insert  into the definition of the partition function , a summation of all the integrals ,each  evaluated on an equivalence class of paths for particles  and field configurations for fields. For Yang Mills fields the partition function requires a sum over the topological micropartitions and functional Action integrals.

$(3)$ QUANTUM STATISTICAL MECHANICS WITH SYMMETRY AND BROKEN SYMMETRY

There are physical systems with symmetry groups whose generators commute with the Hamiltonian. The Casimir operators of the group are  used to obtain the representations of the groups ,  and this gives rise to a degenerate energy spectrum. The partition function , sum over the possible configurations , has to include the degeneracy or multiplicity in the form of the dimension of the representation and the trace over the characters of the group representation.

 When the  symmetry group changes , or a subgroup becomes the symmetry group , or a subgroup of symmetry changes to another subgroup , or the set of operators commuting with the Hamiltonian changes due to terms in the Hamiltonian ;  then the change in the partition function gives rise to a phase transition characteristic of the broken symmetry.

 A quantum mechanical system with a non degenerate energy spectrum is considered ergodic. In the presence of any symmetry group this condition is not obtained. Hence a partial ergodicity and broken ergodicity condition occurs.There can be micropartitions of the density matrix into degenerate and non degenerate sectors. This can be done with partitioned matrices.

 The definition of the partition function involves taking a trace over these sectors . For the degenerate case the representation dependent terms and for the non degenerate case the usual ergodic measure is used. For symmetry group $ G $ and subgroups  $ H_1 $ and $ H_2 $ the quotient $ G/H_1 $ and $ G/H_2 $ is taken each with its own multiplicity index.

Consider the Casimir operators for the groups as $ { C_g}$ and $, {C_h} $ with the dimension of the finite dimensional representation as $d_g$ snd $d_h$ ; which is the degeneracy or multiplicity factor ,or  dimension of the block diagonal form of the matrix representation . The trace , or sum of characters associated with the invariant Casimir operator in this representation  is $ c_g $ and $c_h$ respectively. The Casimir operators commuting with the Hamiltonian give a set of conserved quantities and therefore enter the partition function along with their associated chemical potential. In the usual case of particle number being constant the standard term $\beta \mu N $ occurs  along with $-\beta H $in  the exponent.

 The quantities $d_g,d_h $ and $ c_g ,c_h$ are to be included in the definition of the partition function and density matrix.

The partition function is 

 $ \sum d_g Exp(-\beta E_n) Exp(\beta \sum_i  \mu_ic_{gi})$

 for the group $G $ acting on the system. Similarly there would be a definition for the subgroups. This generalises the usual Trace definition .

$ Z(\beta) = Trace ( Exp( -\beta H)) $  for the Hamiltonian $H$,  which has eigenvalues  $ E_n $. 

For the quantum ergodic hypothesis to be generalised, the Trace can be done on the disjoint representation over the degenerate and the non degenerate parts of the spectrum. Thus $ Z(\beta) = ( Tr _{deg} + Tr_{nondeg}) (Exp(-\beta H) $. Hence the degenerate part sum for the trace is given as above and for the nondegenerate case it will be with the factor $d_g$ of multiplicity set to one.

 For symmetry change or symmetry breaking , the group $G $ is replaced by the subgroups $H_1$ and $H_2$. Then for each case the partition function is evaluated. A phase transition associated with this symmetry change is obtained by using the reduced symmetry partition functions for doing statistical mechanics. The spectrum of the Hamiltonian is assumed to be solvable with these symmetry operations.

The statistical averages are found from the definition of the probability density

 $ \rho  = Exp(-\beta H) Exp(\beta\sum_i\mu_ic_{gi})/Z(\beta) $

$ < A > = Trace (\rho A)/Trace(\rho) $

The partition function  $ \sum_J (2J + 1) Exp(-J(J+1)\beta  h^2/8\pi^2 I) $ is a known  example ,with the degeneracy or multiplicity  factor $ (2J + 1) $

 and the  degenarate energy spectrum is $ (J(J+1)h^2/8\pi^2I)$. This is an example of the representation of the continuous rotation group in the sum for partition function.

For Fermi and Bose gases the partition function has the factor

 $ [ exp(\beta( E_n - \mu)) +\alpha]^{-1} $. 

Where $\alpha $ is $+1$ for Fermions and $-1$ for Bosons and zero for classical gas and $ \mu$ is the chemical potential.

The quantum Liouville equation whose equilibrium solutions these density matrices are expected to be , have partial ergodicity in the sense of the presence of both degenerate and non degenerate spectrum for the Hamiltonian on the micro partitions. It was possible to write a partition function for a mixed representation.

What is the statistical thermodynamics of such types of symmetry group  change  phase transitions. In condensed matter systems the symmetry change occurs in structural phase transitions. Can a system in which  order parameter associated with two subgroups occur be described by this approach. How does the description of the usual first and second order phase transitions due to a critical value of temperature be understood in this framework. Does the nature of the system at the critical point have a group theoretic and topological origin or explanation. These are some questions that have been discussed in various research papers.

As an example consider the collection of  anisotropic oscillators and rotors. The subgroups of the rotation group can be used with their representations and the energy spectrum is known. A degenerate and non degenerate component can be found as a special case . The use of the partition function above could enable symmetry breaking transitions to be defined for this system when a reduced symmetry is obtained in terms of the coefficients in the Hamiltonian.

Finite dimensional irreducible representations of finite  and compact groups  can be used . Modulii spaces arising from the Lattice group modulo such a finite, compact group, such as $ R^2/Z_2 $ and polyhedral groups , space and point groups are widely used in condensed matter physics partition functions to evaluate thermodynamic quantities and phase transitions require the concept of micro partitions introduced here.

$(4)$ QUANTUM STATISTICAL MECHANICS WITH PURE AND ENTANGLED STATES DENSITY MATRICES

The Hilbert space of the N particle system is a n tuplet vector space. In the case that this is completely factorisable as a direct product of one particle Hilbert spaces, the general state is a linear combination of the  tensor product of one particle states. Interactions of two body and three and higher number types can introduce factorisation into reduced spaces of bipartite and tripartite and higher states. Then the Schmidt decomposition for the bipartite case gives a condition for entanglement.

The  $ H^N $ is $ H_1\otimes H_2 \otimes-----\otimes H_N $ ; the total Hilbert space.

For the bipartite case the $ \sum_{ij}\mid i>\mid j> p_{ij} $ gives a linear combination of the  tensor product of two one particle states . This gives a Schmidt decomposition when the $p_{ij} $ are real numbers between zero and one.  The trace condition requires $\sum p_{ii} $ to be equal to one. The entangled states are the more general case here.$ \sum_{ij} p_{ij} ^2 \leq 1 $.

The density matrix formed for one particle states as the outer product $\mid n><m \mid = \rho_{nm} $ satisfies a condition $ \rho = \rho^2 $ and Trace $ \rho $ = Trace $ \rho^2 $ for pure states and Trace $\rho $  $>$ Trace $\rho^2 $ for mixed states.  The states could be taken as the energy basis ; then the  $exp(-\beta H)$; $ H_{nm}$ and  $ Exp(-\beta H_{nm}) $ the operator has the usual representation .  A general combination $  \sum c_{nm} \rho_{nm} $ is also possible. 

The outer product for the bipartite case  also gives a density matrix and the same type of condition to obtain the pure and mixed states.

 $\mid n_1>\mid n_2><m_2\mid<m_1\mid = \rho_{nm} $ The trace condition and the expectation value of an operator is found with this density. 

For a two body interaction model the Hamiltonian split as $ H_{tot} =  H_1 + H_2 + H_{12} $ gives a Gibbs operator $Exp(-\beta H_{tot}) $ . The ordered operators represented in the energy basis could be written with $ H_1 $ acting on the first and the $H_2$ acting on the second ; and the $H_{12} $ acting on both particle states. It will be useful to find the special cases with maximally entangled and pure states, and transitions arising , when a general mixed density is used. to obtain the partition function and the statistical mechanics average value of operators with this $\rho$ ; formally define the  Partition function as the Trace $\rho$. Find the Trace $ (\rho A) $ and take the ratio Trace$(\rho A) $/Trace $\rho$ $= <A>$.

As an example the system of two coupled harmonic oscillators for which known basis functions exist could be tried. This could be done with the transformation to normal modes and quantising them ,as well as in the coupled coordinates in which the three parts of the Hamiltonian can be found and the basis of the single oscillators used to obtain the direct product states. With these states the density matrix is written and the partition function found by taking the trace of  $ Exp(-\beta H) $,using the energy basis. The statistical averages can be found. Special states can be constructed for entanglement and pure cases and the transition studied by finding the expectation values of the operators, such as the Hamiltonian.

For the tripartite and higher case this decomposition may not exist or may not be unique. The density matrices formed out of outer products of these multi partite states generalise the one particle density matrices of the usual quantum statistical mechanics. The true representation of the quantum statistical mechanics is the collection of such density matrices and the operators acting on them. When three body and many body interactions are important , the tripartite and multipartite basis will be needed. This case could be worked out on the same lines as the bipartite one , but that may require additional limitations.

In diatomic molecules the bipartite case is useful. The triatomic and polyatomic molecules are also treated with two body interactions due to weak coupling. In nuclear and particle physics the three and many body interactions become important and the tripartite and multipartite case is neccessary. For a two quark in a meson, and three quark in a hadron example the possibility of a Schmidt decomposition on the quark states for mesons is available.  For the tripartite case in the three quark interaction in the hadron , it will be interesting if the GHZ and W type of states are possible ; these are unique in being expressed as the direct product one particle states.

 The maximally entangled bipartite states are obtained from $\mid 00>,\mid 11>, \mid 10>, \mid 01> $ by  a correct linear combination , and the tripartite states are obtained from $\mid 000> ,\mid 111> , \mid 100> ,\mid 010> , \mid 001> . $    With these states used to calculate the density matrices and then the partition function ,the entanglement dependence is tested. The discrete representation for the `` spin or other''quantum properties and  the corresponding Energy eigenvalues , are thus included in the partition function Trace sum. The interpretation of the transition arising from the types of basis used as phase transition for the micro partitions is suggested.

 The partition function is the trace of the density matrix, usually expressed in the energy eigenstate basis for the Gibbs form. In the N particle system Hamiltonian the possibility of having mostly a two body contribution to the potential energy will permit factorisation into a bipartite basis and hence a clear definition of pure, mixed and entangled states.

 Using this assumption which is valid in a weak coupling low density system; the basic partition function can be written . The outer product of bipartite states,  which are themselves written as a tensor product of one particle states, is made. The coeffecients of the linear combination of such terms are subject to the Schmidt decomposition condition.

 The pure and entangled special cases arising from the general mixed state density matrix ,  can be considered as micro partitions on the space of states. The direct sum of the subspaces of these types of states gives the total space of the bipartite distributions.

 Thus the N particle Hilbert space is  now reduced to a direct sum of the pure, entangled and mixed sectors of bipartite basis densities. The trace operation on this direct sum is a direct sum of the traces on each sector. The partition function is now written in terms of these micropartitions. The true density for statistical mechanics of the N particle system for obtaining the averages ,  is the ratio of the bipartite density and the partition function which is the trace of that density.

 The trace of the operators that are represented on this basis density  divided by the trace of the density is the average value of the operator. The phase transitions that result from the changes in this decomposition into the pure, entangled and mixed sectors are the interesting ones .

 One special case could be to evaluate the transition from the maximally entangled basis to the pure state basis. When is this a smooth change , when is it a sharp change. The approach to equilibrium creates loss of information or correlations and due to decoherence of the entangled states this leads to the pure basis.  For a free particle system this is expected. The phase space version of quantum mechanics and the Wigner distributions for the entanglement could be used for a general framework.

 For a system of particles in a common potential or with two body long range interactions the entanglement is not expected to decay and in this case the phase transition will be controlled by the interaction. The statistical mechanics of entanglement is still an open subject. For mixed state density matrices which are sparse matrices, there are computable effects , but the general case does not lead to a testable result that will help create a formalism.

 CONCLUSION

The foundations of statistical mechanics have been redefined over a period of more than a century. As more features of classical and quantum mechanics underlying statistical mechanics were known and as the probabilistic methods were developed. 

The application of statistical mechanics as a method to a variety of systems in condensed matter, high energy physics and gravitation and to equilibrium and non equilibrium processes with thermodynamic properties have also required adaptation of the fundamentals of statistical mechanics.

 The developments in a variety of these subjects in the past few decades neccesitate a fresh consideration of the foundations and the applications.

 One example of the former , that is of the foundations,  are the four topics dealt with above ; that did  not arise in the classic works of the foundations of statistical mechanics,  as it was known thirty years ago ,  but have occured later. These poossibilities have already made significant impact in physics and hence require inclusion at the foundations of the formalism of statistical mechanics ; and a small beginning is made here. 

An example of the latter , that is of application,  that  is the  development of exclusion statistics and anyons that was surprisingly missed earlier; these have been extensively worked out in the past decade. When basic texts of statistical mechanics are written with these newer properties included , the redefined foundations of statistical mechanics will be available.

The concept of micro partitions was introduced to enable dealing with  underlying structures that have their roots in classical and quantum mechanics. This creates additional possibilities for phase transitions and requires a more general formalism of statistical mechanics based on redefining the partition functions.

The usual average over ensembles of independent configurations , energy levels and numbers of particles is generalised due to the averaging over the micro-partitions or components.

 ACKNOWLEDGEMENT

I thank the Institute of Mathematical Sciences , Chennai ,  and especially its Director ,for  facilities and supporting my visits to the Institute. 

The new possibilities in mathematical physics and their inclusion in fundamental physics , was the objective of my work. This was motivated by my profession as a physics teacher , with research interests in theoretical physics.

 This sixth eprint in www.arxiv.org will also be accessible to those who are interested in these themes.

 REFERENCES

(1) Statistical Mechanics , G. Morandi, F.Napoli, E.Ercolessi, World Scientific ,2001

(2) Quantum computation and quantum information, M. Nielsen and I. Chuang , Cambridge university press,2000

(3) Group theory and physics, S.Sternberg , Cambridge University press ,1994

(4) C.Grosche and F Steiner, Handbook of Feynman Path Integrals, Springer Texts of Modern physics, vol 145, 1998

(5) R.K.Pathria Statistical Mechanics, Butterworth and Heinman .1996

(6) L. E. Reichl,Statistical Physics,Arnold, Unwin 1983

(7) A. B. Balentekin, Partition functions in Statistical Mechanics ,symmetric functions and group representations PRE 64(2001) 066105, hep-th/0007161

(8) S.T.Hong hep-th/0104149 in www.arxiv.org

(9) H.J.Schmidt, J.Sohnack cond-mat/0104293 ,

(10) Robert Shrock cond-mat/9908323 and hep-th/9811166

(11) K.Huang , Quantum field Theory , Wiley Interscience,1998

(12) N.D.Hari Dass, S.Kalyana Rama, B.Sathiapalan ,cond-mat/0112439,Int Jnl of Mod Phys A 18,2003 ,p2957-2994

(13) Ajay Patwardhan, quant-ph/0211039,0211041,0305150  , hep-th/0406049   

(14) M.Nakahara , Geometry, Topology and Physics, Institute of Physics publication 1987

\end{document}